\documentclass[11pt]{article}
\usepackage[utf8]{inputenc}
\usepackage[tmargin=1in,bmargin=1in,lmargin=1in,rmargin=1in]{geometry}
\usepackage{graphicx}
\usepackage{float}
\usepackage{amssymb}
\usepackage{adjustbox}
\usepackage{epstopdf}
\usepackage{enumitem}
\usepackage{amsmath}
\usepackage{biblatex}
\usepackage{textcomp}
\usepackage{chngcntr}
\usepackage{comment}
\usepackage{caption}
\usepackage{subcaption}
\usepackage[T1]{fontenc}
\usepackage{bm}
\usepackage{float}

\usepackage{authblk}
\usepackage{multirow}
\usepackage{multicol}
\usepackage{comment}
\usepackage{placeins}
\usepackage{ragged2e}
\allowdisplaybreaks

\usepackage{hyperref}
\hypersetup{
    colorlinks=true,
    linkcolor=blue,
    filecolor=magenta,      
    urlcolor=cyan,
    citecolor=blue,
}
\title{Implications of regional variations in climate change vulnerability and mitigation behaviour for social-climate dynamics}
 \author[1,2,*]{Amrita Punnavajhala}
 \author[3]{Timothy M. Lenton}
 \author[1]{Chris T. Bauch}
 \author[2]{Madhur Anand}

\affil[1]{Department of Applied Mathematics, University of Waterloo, Waterloo, Ontario, Canada}
\affil[2]{School of Environmental Sciences, University of Guelph, Guelph, Ontario, Canada}
\affil[3]{Global Systems Institute, Faculty of Environment, Science and Economy, University of Exeter, Exeter, United Kingdom}
\affil[*]{Corresponding author: apunnava@uwaterloo.ca}
\begin{document}

\maketitle
\begin{justify}

How regional heterogeneity in social and cultural processes drive--and respond to--climate dynamics is little studied.  Here we present a coupled social-climate model stratified across five world regions and parameterized with geophysical, economic and social survey data.  We find that support for mitigation evolves in a highly variable fashion across regions, according to socio-economics, climate vulnerability, and feedback from changing temperatures. Social learning and social norms can amplify existing sentiment about mitigation, leading to better or worse global warming outcomes depending on the region.  Moreover, mitigation in one region, as mediated by temperature dynamics, can influence other regions to act, or just sit back, thus driving cross-regional heterogeneity in mitigation opinions. The peak temperature anomaly varies by several degrees Celsius depending on how these interactions unfold. Our model exemplifies a framework for studying how global geophysical processes interact with population-scale concerns to determine future sustainability outcomes.

\end{justify}

\newpage
\section*{Introduction}

  Anthropogenic climate change is underway and its myriad and growing effects range from frequent and severe weather events \cite{RN15}, to increased disease transmission \cite{patz2005impact}, as well as conflict, violence and human migration \cite{nyt}. The impacts will extend beyond humans to many varieties of ecosystems and their biodiversity \cite{RN15,kaplan2006arctic, meyer2024temporal, trisos2020projected}. It will take a combination of both mitigation and adaptation to address the climate change crisis \cite{dessler2021introduction}. 

The Intergovernmental Panel on Climate Change (IPCC), whose reports are used to inform policy worldwide, has curated results from a substantial number of models studying climate change \cite{ipcc_reports}. These models are nuanced in their ability to account for a variety of socio-economic and geo-political scenarios, including projections of population, income, technology costs and future government policies (most recently in the form of the Shared Socioeconomic Pathways \cite{riahi2017shared}). On the other hand, these models lack mechanistic representations of human behavioural dynamics, assuming that humans are homogeneous economic agents \cite{weyant2017some, schwanitz2013evaluating}. The study of climate change mitigation, however, benefits from an understanding of human behaviour and especially social factors \cite{nielsen2024realizing}. Individuals directly impact mitigation efforts through the choices they make as consumers, such as lifestyle choices that reduce carbon footprints, choosing to eat less meat \cite{stehfest2009climate}, using public transport \cite{chapman2007transport}, or long-term decisions, often involving substantial financial outlays, such as installing solar panels \cite{hu2016impact}. Individuals also have the potential to collectively influence policy through their votes or even through protests or other large-scale social movements (for instance, the `Fridays for Future' campaign \cite{thiri2022social}). Similarly, it has been found that when injunctive norms (social pressure) that promote sustainability clash with negative descriptive norms (the perception that others do not follow the behaviour), the effectiveness of injunctive norms diminishes \cite{ge2020solve}.   

Although economic costs are an important driver of the adoption of clean energy, social processes are increasingly recognized as equally important, and more importantly, the success of policies can depend on the interaction between economic and social factors \cite{kinzig2013social}. Policy measures often require the support of individuals to be fully effective, as demonstrated during the COVID-19 pandemic, where control depended critically on individuals' willingness to comply with government regulations \cite{chan2020confidence, gelfand2021relationship}. Moreover, social norms significantly influence sustainable behaviors \cite{yamin2019using}, even to the point where cost-saving measures might not be adopted if they conflict with personal or perceived social norms \cite{niu2023role}. For example, in business travel culture, the perception of professional expectations and status associated with business air travel--even for short flights--can prevent the use of cleaner forms of transport \cite{arrondo2022environmental}. Similarly, the consumption of costly beef products instead of less expensive and healthier vegetarian food is partly driven by entrenched social and cultural forces \cite{sanchez2019consumer}. Hence, social processes can prevent the adoption of sustainable behaviour, even when that behaviour is less economically costly. 

Recent years have seen the emergence of coupled social-climate models that account for two-way feedback between such social processes and climate change. Early models found that behavioral and physical uncertainties are of comparable magnitude in influencing climate change \cite{beckage2018linking, bury2019charting}. Social climate models are an example of socio-ecological or coupled human-and-natural system models for other systems that have been studied longer \cite{innes2013impact, richter2015profit, thampi2018socio, henderson2016alternative}. These models have shown how accounting for social dynamics can result in novel system states that are not found in ecological or social models in isolation from each other \cite{schluter2016robustness, lade2013regime}, suggesting a valuable role for coupled social-climate models. 

Discussions on coupled social-climate models have presented new directions for future research \cite{schluter2019potential, calvin2018integrated, farahbakhsh2022modelling,nielsen2020improving, donges2021taxonomies, shults2021artificial}, as well as new types of models \cite{donges2020earth, muller2021anticipation,moore2022determinants,bechthold2024social}. There is also a call for consolidating and creating ensembles of these models to improve robustness \cite{donges2021taxonomies}. Until now, social-climate models have used relatively stylized descriptions of human behavior in both homogeneous \cite{bury2019charting, beckage2018linking} and heterogeneous populations \cite{menard2021conflicts}. Modelling social dynamics across diverse populations allows for a more realistic representation of human behaviour \cite{beckage2022incorporating} and also offers more nuanced insights into social-climate dynamics, compared to models with homogeneous populations \cite{menard2021conflicts}.  

We construct, for the first time, a coupled social-climate model representing the world population broken down into the five regions defined in the Shared Socioeconomic Pathways \cite{bauer2017shared}. Through this, we are able to capture regional differences in some of the factors that influence mitigation support. These include diverse social and cultural settings \cite{andre2024globally,eriksson2021perceptions, simpson2021climate,spektor2023climate, niamir2020assessing}, differences in vulnerability to climate change caused by variations in exposure to extreme weather \cite{kotz2024economic,JRC129896, xu2020future, ricke2014natural}, and socioeconomic circumstances \cite{kotz2024economic}. Understanding mitigation at the regional level is also useful because global agreements on mitigation targets are influenced by interactions at this scale \cite{karatayev2021well, vasconcelos2014climate}. Our treatment of social dynamics uses the replicator equation from evolutionary game theory \cite{cressman1995evolutionary, traulsen2023individual, arefin2024coupling}, accounting for perceived climate change severity, learning processes, injunctive social norms \cite{bonan2020interaction} and economic costs. We use our model to explore the evolution of support for mitigation across multiple regions, and its impact on the global temperature anomaly.  Our objective is to uncover aspects of social-climate dynamics that are unique to models with regional structure and provide a framework for developing geographically structured coupled social-climate models. 

\section*{Modelling framework}
\begin{figure}
    \centering
    \includegraphics[width = \linewidth]{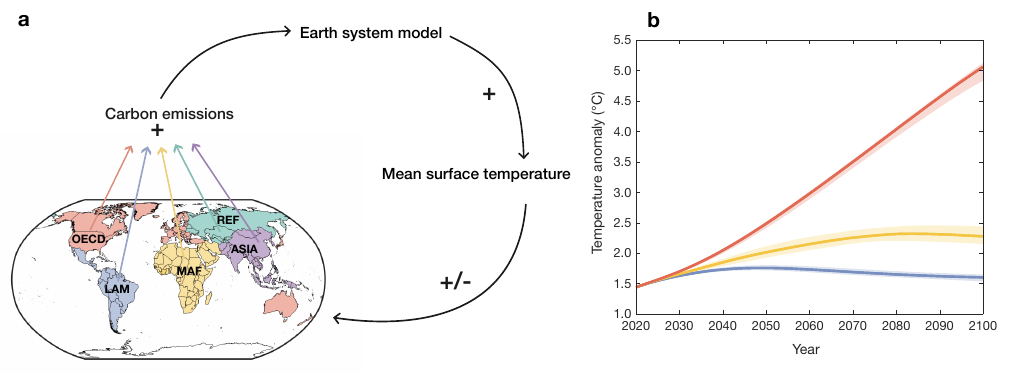}
    \caption{\textbf{Model diagram and calibrated scenarios: baseline, best case and worst case.} (a) In our model, carbon emissions from each region are summed and fed into the earth system model which computes the temperature, which then goes on to affect mitigation in each region. (b) Model dynamics generate the anomaly in the global mean surface temperature (in $^\circ$C, relative to pre-industrial levels), under the best case (blue), worst case (red) and baseline (yellow) scenarios. Region-specific social dynamics are not shown here.  The solid line shows median trajectories for the filtered parameter values, with 0.25 and 0.75 quantiles represented by the shading.}
    \label{fig:model_diagram}
\end{figure}

We couple a model of social dynamics with a simplified earth system model (Figure \ref{fig:model_diagram}a). The social model represents dynamic decision-making processes regarding support for climate change mitigation. The model output determines the extent of climate change mitigation, which then determines anthropogenic carbon emissions. These carbon emissions are treated as input to the climate component of the model, which churns out a temperature anomaly (relative to pre-industrial levels). The temperature anomaly influences human perception of the severity of climate change, which once again affects social dynamics and stances on supporting mitigation. Since our goal is not to predict the future with a `black box' model, but rather to gain insights into how social-climate models with regional structure behave, we make simplifying assumptions to facilitate model analysis. 

We assume that social dynamics are specific to five world regions: Asia (ASIA), Latin America (LAM), the Middle-east and Africa (MAF), Organization for Economic Co-operation and Development (OECD) countries and the Reforming Economies of Eastern Europe and the former Soviet Union (REF). These regions are as defined in the Shared Socioeconomic Pathways (SSPs) five region aggregation \cite{bauer2017shared}.  Notably, we do not account for social interactions between regions, though all regions are coupled indirectly through their climate change impacts. Factors extrinsic to the model, such as economic and demographic changes, are based on `middle-of-the-road' SSP 2 projections for which data are available \cite{riahi2017shared}, else, under assumptions closest to those of SSP 2.

In our model, individuals can choose either to support climate change mitigation or not. The proportion of a population supporting mitigation (`mitigators'), $x$, grows if the utility of switching to support mitigation is positive. Individuals interact with each other at the social learning rate, $\kappa$, and through interactions are able to gauge whether they would benefit from switching behaviour. The higher the learning rate, the more a person interacts with others. And, the higher the net utility gain, the more likely they are to change behaviour. The net utility in each of the five regions is based on empirical estimates of the net cost to mitigate $(\beta - k)$, the cost of climate change impacts $f(T)$, and the strength of social norms $\delta$, specific to each region.
The net cost associated with mitigation is based on empirical estimates of per capita costs of renewable energy, $\beta(t)$, relative to fossil fuels, $k$, calculated by multiplying renewable energy costs \cite{way2022empirically} with per capita energy demand projections \cite{bauer2017shared}, and then subtracting from that the costs of fossil fuels, which are projected to be relatively constant \cite{way2022empirically}. We assume that the cost $f(T)$ associated with climate change impacts is a function of $T(t)$, the current anomaly in the global mean surface temperature. The functional form for $f(T)$ is based on estimated national damages to GDP per capita under RCP 8.5 warming \cite{kotz2024economic}. To estimate the strength of social norms, $\delta(t)$, as well as the initial fraction of mitigation supporters in each region, $x_0$, we use survey data on support for fighting global warming \cite{andre2024globally}.

Parameter estimates are multiplied by scaling factors and then re-arranged to non-dimensionalize the model equations, thereby reducing the number of parameters to avoid overfitting. We use Approximate Bayesian Computation (ABC) \cite{thommes} to estimate the scaling for the initial fraction of mitigators, $\hat{x}_0$, the rescaled social learning rate, $\kappa_0$, the cost of climate change impacts, relative to mitigation costs (`relative climate impact cost'), $c_0$, and the strength of social norms relative to mitigation costs (`relative norm strength'), $\delta_0$. We calibrated these parameters to best case, baseline, and worst case scenarios. For our baseline scenario, the modelled carbon emissions trajectory was calibrated to the RCP 4.5 scenario, thereby generating plausible ranges for the remaining parameters. For our best case scenario, we similarly calibrated the model to the RCP 2.6 scenario, and for the worst case scenario we used RCP 8.5 \cite{pachauri2014climate}. (Supplementary Information, Table \ref*{tab:params} for calibrated parameter point estimates and ranges). To represent climate dynamics in our coupled model, we adapt an existing earth-system model \cite{lenton}. The model describes the carbon cycle, representing dynamics in the atmosphere, ocean, vegetation and soil, as well as changes in the global mean surface temperature, as influenced by the volume of atmospheric carbon.

\section*{Results}

Our modelling results predict significant regional heterogeneity, with population opinion on mitigation progressing in a highly variable fashion across regions and scenarios due to differing responses to temperature anomalies and socio-economic circumstances. Social learning can have divergent impacts on public support for mitigation, depending on perceived climate damages and social norm strength.  Moreover, social dynamics in one region can affect other regions through the intermediary of temperature feedbacks, allowing the possibility that one region can get a `free ride' on the mitigation efforts of other regions. We describe these results in more depth in the following subsections. 

\begin{figure}
    \centering
    \includegraphics[width = \linewidth]{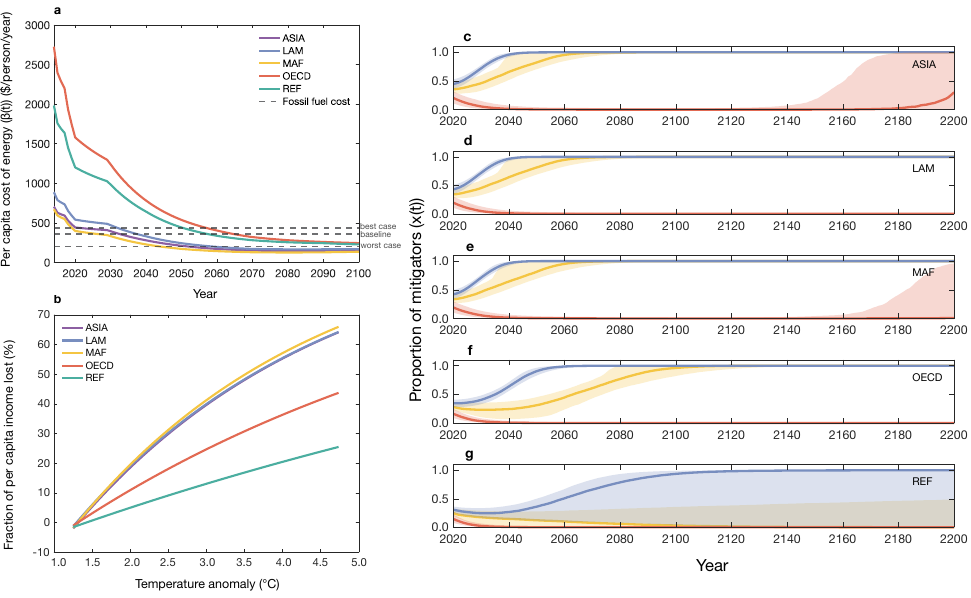}
    \caption{\textbf{Cost parameters drive mitigation support most strongly in ASIA, LAM and MAF.} (a) Declining per capita costs of renewables and calibrated per capita fossil fuel costs under best case, baseline and worst case scenarios. (b) Per capita percentage loss in income as a function of temperature anomaly. (c-g) Mitigation support over time for the best case (blue), baseline (yellow) and worst case (red) scenarios, with solid lines showing median trajectories for filtered parameters and shadows outlining 0.25 and 0.75 quantiles.}
    \label{fig:x_params}
\end{figure}

\paragraph{Characterizing mitigation in each region.}
\begin{table}[]
    \centering
\begin{tabular}{|l|c|c|}
\hline
Region     & Initial fraction of mitigators ($x_0$) & Strength of social norms ($\delta$) \\
\hline
 ASIA    & 0.76 & 0.66 \\
 LAM    &  0.73 & 0.67\\
 MAF    &  0.72 & 0.60\\
 OECD &  0.60  & 0.58 \\
 REF &   0.54  & 0.56\\
 \hline
\end{tabular}
    \caption{\textbf{Estimated initial mitigation support and social norm strength.} See Methods for details. The strength of norms is a number between 0 and 1, with larger values corresponding to stronger norms.}
    \label{x0_delta_table}
\end{table}

We first give an overview of mitigation regimes in each of the five regions. Economic projections show how mitigation costs decline with time in all regions, driven by projected decreases in the costs of renewable energy \cite{way2022empirically}; the larger the per capita energy consumption for a region, the more expensive they find the transition to renewables (Figure \ref{fig:x_params}a). The cost of renewables eventually falls below that of fossil fuels, so the net cost turns into a net gain from mitigation.  However, it is possible that majority-enforcing social norms can `lock in' a population at low levels of support for mitigation, even after the net cost of renewables turns negative. The OECD and REF, compared to other regions, have both higher initial mitigation costs (Figure \ref{fig:x_params}a) and lower perceived relative costs of climate change impacts (Figure \ref{fig:x_params}b), suggesting that mitigation in these two regions is less attractive than it is in ASIA, LAM and MAF. 

Next we consider projected mitigation dynamics in each region under three scenarios: baseline, best case and worst case. Mitigation support in ASIA, LAM and MAF (Figure \ref{fig:x_params}c-e) outpaces that in the OECD and REF (Figure \ref{fig:x_params}f,g) in the best case and baseline scenarios. Under baseline conditions, mitigation has majority support in ASIA, LAM and MAF soon after 2030 in contrast to the OECD, which reaches majority in about 2060, and the REF, where support dwindles to near zero by 2090. In the worst case, support dies out in all regions. We also inferred the strength of social norms from survey data \cite{andre2024globally} and found that ASIA and LAM have the strongest norms as well as the largest initial proportion of mitigators (Table \ref{x0_delta_table}).  

\begin{figure}
    \centering
    \includegraphics[width=\linewidth]{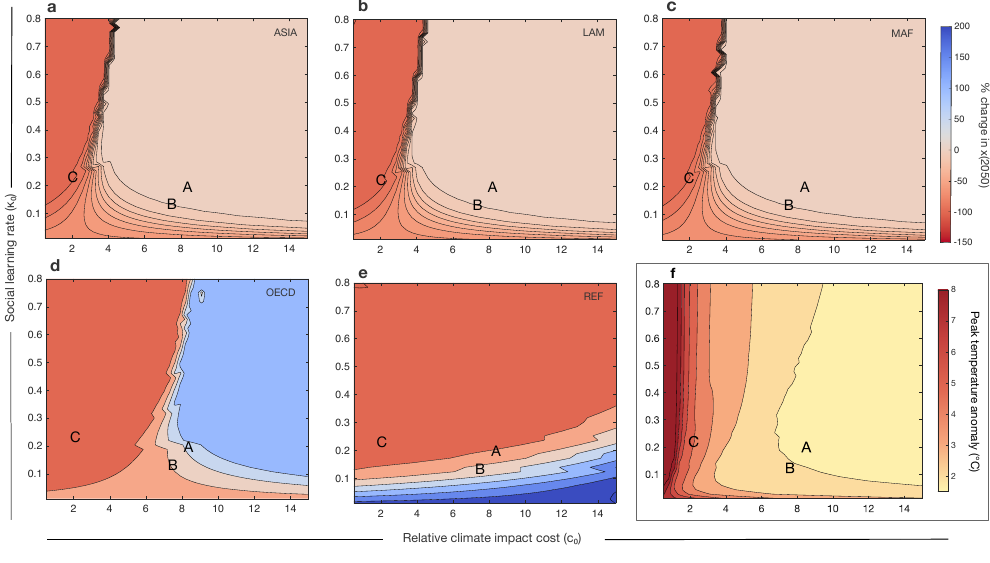}
    \caption{\textbf{Increased social learning can slow mitigation when relative climate impact cost is low.}  Shown is the percent change in the fraction of mitigators in 2050 relative to baseline for various combinations of the learning rate ($\kappa_0$) and relative climate impact cost ($c_0$) in each region varied separately (a - e), and the peak temperature anomaly when learning rates and relative climate impact costs are varied for all regions simultaneously (f). On each plane, `A' marks the best case, `B', baseline, and `C', the worst case parameter combinations.}
    \label{param_planes_k_c0}
\end{figure}

\begin{figure}
    \centering
    \includegraphics[width=\linewidth]{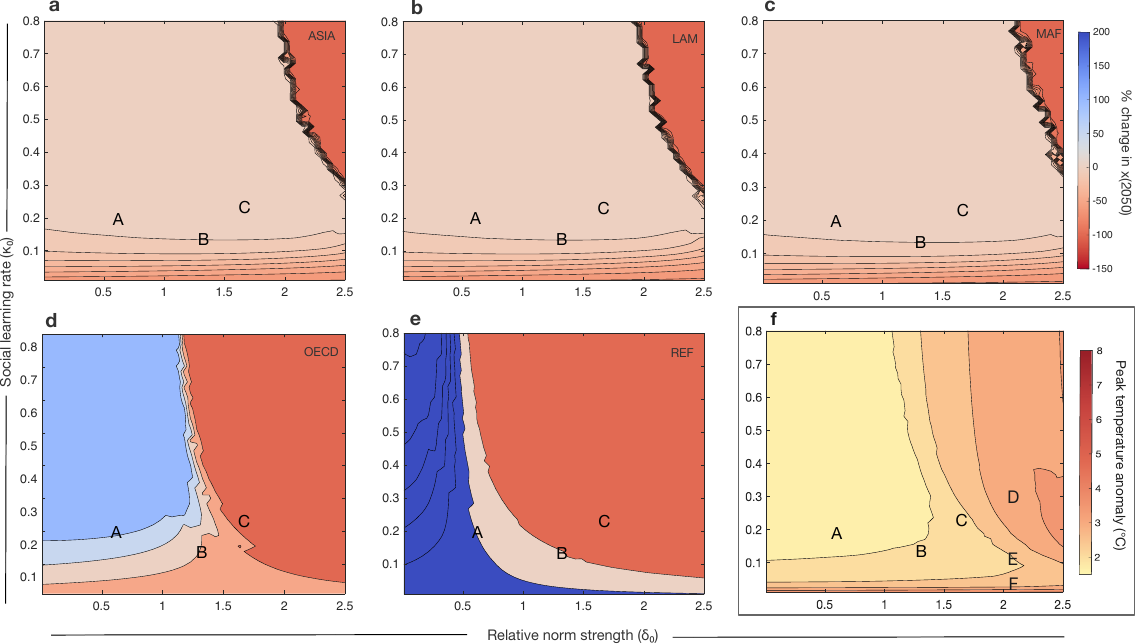}
    \caption{\textbf{Norms and learning rates affect outcomes in OECD and REF more than ASIA, LAM and MAF.} Shown is the percent change in the fraction of mitigators in 2050 relative to baseline for various combinations of the learning rate ($\kappa_0$) and relative norm strength ($\delta_0$) in each region varied separately (a - e), and peak temperature anomaly when learning rates and relative norm strengths are varied for all regions simultaneously (f). On each plane, `A' marks the best case, `B', baseline, and `C', the worst case parameter combinations. The time evolution of opinion and temperature for parameter combinations `D', `E' and `F' in (f) are shown in Figure \ref{fig:temp_feedback}c-e. }
    \label{param_planes_k_d0}
\end{figure}

\paragraph{Social processes affecting mitigation.}
The same social processes can generate very different outcomes in different regions. ASIA, LAM and MAF  are affected in similar ways by changes in the social learning rate $\kappa_0$ and the relative cost of climate impacts $c_0$ (Figure \ref{param_planes_k_c0}a-c, Figure \ref{param_planes_k_d0}a-c), but outcomes in the OECD and REF look quite different (Figure \ref{param_planes_k_c0}d,e, Figure \ref{param_planes_k_d0} d,e)). For instance we see that in ASIA, LAM and MAF there is little room to increase support in 2050 though a drop in relative climate impact costs or increase in relative norm strength, coupled with increased learning rates can slow support by close to 100\%, relative to baseline.  This is because the support for mitigation in these regions is projected to remain robustly high, except for the worst case scenario. On the other hand, early mitigation support in the OECD and REF shows potential to grow or decline depending on the social learning rate and relative strength of norms. Specifically, increased social learning can scale up support in 2050 by as much as 100\% in the OECD if relative climate impact costs are high (Figures \ref{param_planes_k_c0}d) or norms weak (Figure \ref{param_planes_k_d0}d). In the REF, decreased social learning together with a higher climate impact cost (Figure \ref{param_planes_k_c0}e) can increase support by up to 200\%.

An interesting and novel finding is that social learning can work two ways: increased social interactions can either drive up support for mitigation, or cause it to collapse, depending on prevailing levels of relative climate impact costs and norm strengths. The higher the learning rate, the faster the strategy with higher utility spreads. In ASIA, for example, baseline opinion (point `B' in Figures \ref{param_planes_k_c0}a, \ref{param_planes_k_d0}a) favours mitigation strongly enough that a higher learning rate leads to a better mitigation outcome, while in the REF, a lower social learning rate helps slow the spread of negative sentiment about mitigation (starting from baseline point `B' in Figure \ref{param_planes_k_c0}e). We also see that support in ASIA, LAM and MAF (Figures \ref{param_planes_k_c0}a-c, \ref{param_planes_k_d0}a-c) is less easily perturbed by parameter changes compared to the OECD and REF (Figures \ref{param_planes_k_c0}d,e, \ref{param_planes_k_d0}d,e). In all regions, a high level of initial support can help prevent future declines caused by low climate impact costs or strong norms (Supplementary Information, Figure \ref{fig:SI_param_planes_4_6} f-o).  We also observe that higher climate cost or weaker social norms both support mitigation opinion, but parameter planes for both parameters indicate some regimes of rapid change in opinion, but other regimes of diminishing returns (Supplementary Information, Figure \ref{fig:param_planes_3_SI}a-e).

\paragraph{Feedback between social and climate systems.}

\begin{figure}
    \centering
    \includegraphics[width=\linewidth]{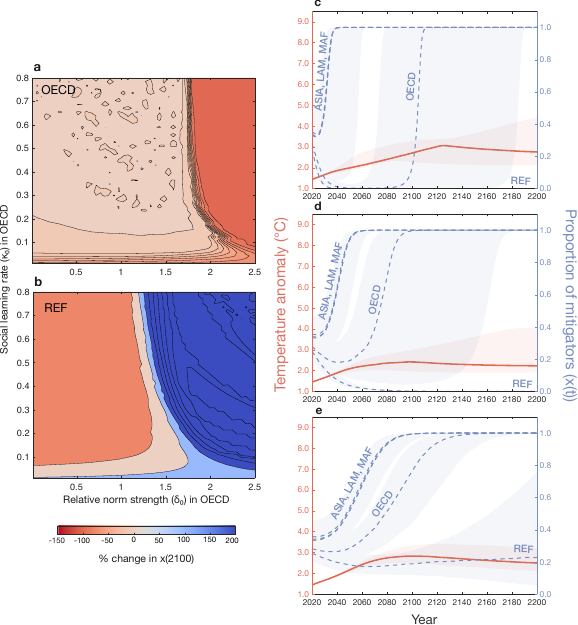}
    \caption{\textbf{Temperature response and subsequent feedback can lead to free riding.} Percent change in mitigation support in 2100 in the OECD (a) and REF (b), relative to baseline, as a function of  changing the learning rate and relative norm strength in the OECD. The time evolution of mitigation support and temperature for marked points `D', `E' and `F' in Figure \ref{param_planes_k_d0}f are shown in (c)-(e), respectively, with temperature in red (left y-axis) and mitigation support in blue (right y-axis), and shadows outlining 0.25 and 0.75 quantiles.}
    \label{fig:temp_feedback}
\end{figure}

The social processes in our model are not just a function of socio-economic differences between isolated regions.  Rather, we find that the social dynamics in each region influence the dynamics of other regions, through the mediation of the global temperature anomaly.  Modifying the social parameters in one region can affect mitigation support in other regions, despite there being no direct coupling between them.  For instance, we see that increasing the strength of norms in the OECD can reduce mitigation support in 2100 by close to 100\% (Figure \ref{fig:temp_feedback}a). However, this has the side effect of increasing mitigation support in REF by almost 200\% (Figure \ref{fig:temp_feedback}b), as global warming is felt more severely because of the OECD's inaction, and REF populations decide to respond more strongly. On the other hand, we see that if norms are weak in the OECD, keeping mitigation close to baseline levels, support in the REF can drop by about 75\%, demonstrating free riding. 

We also find that the learning rate can have a non-monotonic effect on temperature outcomes. For certain values of the relative norm strength, there is a middle range of learning rates for which the peak global temperature anomaly is at its lowest value, and higher or lower learning rates outside of this range lead to worse temperature outcomes (Figure \ref{param_planes_k_d0}f). Strong norms cause mitigation support to dip initially in all regions, resulting in a higher global temperature anomaly. But, as temperature effects begin to be felt, opinion on mitigation starts to reverse. The time scale at which this reversal takes place in different regions is what leads to different temperature outcomes (Figure \ref{fig:temp_feedback}c-e). If the learning rate is high ($\kappa_0 = 0.3$, $\delta_0$ = 2.2), the resurgence of mitigation in ASIA, LAM and MAF is very fast (before 2030), pushing the temperature trajectory down. This lulls the OECD into a false sense of security, delaying its response beyond 2080, and the temperature anomaly eventually hits a peak value of almost 3.1$^\circ$C (Figure \ref{fig:temp_feedback}c). On the other hand, if the learning rate is low ($\kappa_0 = 0.03$, $\delta_0 = 2.2$), mitigation support in ASIA is slow to spread even as global warming effects are felt. ASIA's slow response drives the OECD to pro-mitigation sentiment earlier than in the high-learning case, but the peak temperature of about 2.8$^\circ$C (Figure \ref{fig:temp_feedback}e) is still higher than the value of 2.5$^\circ$C observed in the moderate-learning case ($\delta_0 = 2.2$, $\kappa_0 = 0.1$) (Figure \ref{fig:temp_feedback}d). Through this we see that regions like the OECD and REF that are inherently less disposed to mitigation show the potential to free-ride on mitigation efforts of other regions. We also note that a lower learning rate $(\kappa_0 = 0.03)$ results in a more unified mitigation response across regions. As we increase the learning rate, there is a wider gap between what is happening in ASIA, LAM and MAF, and the OECD and REF (Figure \ref{fig:temp_feedback}c-e).

Strong social norms can counteract the benefits of increased relative climate impact costs on peak temperature, so the amplification of feedback from within the social system (via $\delta_0$) can have almost as much of an impact as feedback from the climate system (via $c_0$) on global warming outcomes (Figure \ref{fig:norms_temp}a). If initial support for mitigation is not very strong ($x<0.5$), norms make it difficult for pro-mitigation sentiment to spread. This delay in early mitigation action can cause severe warming outcomes (Supplementary Information, compare Figure \ref{fig:norms_temp}b, c), that cannot be undone by the benefits norms bring once the majority supports mitigation ($x >0.5$). One way to avoid such outcomes is by stepping up the relative cost of climate impacts (Figure \ref{fig:norms_temp}a).

\FloatBarrier
\section*{Discussion}

Much previous research on climate change mitigation makes an implicit assumption that socio-economic differences between populations are adequate to explain observed differences in support for mitigation across populations, without the need to invoke feedback from temperature dynamics on behaviour or interactions between populations.  On the other hand, coupled social-climate models in the absence of socio-economic heterogeneity suggest that social processes should always have the same impact on mitigation behaviour, in different populations.  Our regional-level social-climate model parameterized with region-specific socio-economic data suggests a third perspective: the same social processes can generate very different outcomes in different regions, and interactions between these regions are mediated by the dynamics of the global temperature anomaly. 

Our coupled social-climate model with regional structure predicts that mitigation progresses more rapidly in ASIA, LAM and MAF than it does in the OECD and REF, which is consistent with empirical research \cite{bergquist2019experiencing, demski2017experience}. We also find that mitigation support in ASIA, LAM and MAF is stable across a broader range of parameter values than in the OECD and REF. On the other hand, ASIA, LAM and MAF could face challenges to implementing climate change mitigation measures that are not explicitly captured in our model, such as less access to clean technologies \cite{jakob2014feasible, pfeiffer2013explaining, sovacool2022equity,ipcc6arsynth}, or trade-offs between mitigation and development \cite{jakob2014climate, tavoni2015post}.  

Our findings also suggest that increased public discourse can aid the spread of opinions in favour of climate change mitigation \cite{wang2020communicating} but only when climate change damages are appropriately emphasized, and anti-mitigation social norms are not too firmly entrenched.  More broadly, a focus on the content of information accessible to the public could help prevent the transmission of misinformation \cite{osborne2022science, farrell2019evidence} and also decrease pessimism about climate action \cite{andre2024globally, cherry2024discourses}, both of which could help with timely mitigation.  But even then, our model shows that some populations may be very resistant to change. In a related vein, we see that mitigation trends in one region can influence mitigation in other regions purely through feedback via global warming. This could result in regions free riding or, alternately, picking up the slack for others' lack of mitigation. This emphasizes the need for binding global-scale climate change treaties \cite{barrett2008climate}, although starting with national or regional level treaties may be a useful stepping stone \cite{karatayev2021well}.

Our work is a first step in representing the world's regional socio-cultural structure in a coupled social-climate modelling framework. However, there are many dimensions that we do not account for. Our assumption that social dynamics are local to each region does not account for the role of smaller scale social groups.  Our region-wide approach fails to account for marginalized, vulnerable groups, such as Indigenous populations in developed regions like the OECD \cite{ipcc6arsynth}. Similarly, we do not include cross-regional connections through flow of information or people \cite{altman2023globalization}. Future work could study the role of cross-border migration attributable to climate change, and its demographic and social impacts \cite{kaczan2020impact}. Many of these limitations could be remedied by building models that represent social dynamics at finer scales, such as national or sub-national.  Likewise, our model implicitly assumes that social processes are sufficient to determine emission trajectories, but this assumes governments that can effectively respond to the average opinion about climate change in their population. Dynamics could change if international negotiations and feedback between different scales of social and/or political dynamics are modelled. Finally, we note that our model parameterisation is conservative, in the sense that we assume a slower decline in the cost of renewables than what has been found in other contemporaneous analyses \cite{nijsse2023momentum}. Despite this, we expect our conclusions would not be qualitatively altered under a faster decline in the cost of renewables, since social norms can still cause a delay in their adoption, and regional heterogeneity in opinions should still emerge due to differences in climate impacts. 

These limitations suggest future directions for social-climate modelling research that could improve their validity and predictive power. The incorporation of social dynamics into integrated assessment models could be especially useful.  The trajectory of model sophistication often increases over time in many fields: the earliest published climate models were simple one-dimensional radiative convective models, for instance \cite{manabe1967thermal}. Social-climate models could exhibit a similar trajectory. However, simple models also have their advantages, such as transparency, ease of analysis, and ability to generate insights. Our regional-level model shows the crucial role played by both social processes and regional heterogeneity in determining climate trajectories. It also exemplifies a framework that could be used to study how global geophysical processes interact with human-scale concerns--economic well-being, equity, and social norms--to determine future sustainability outcomes.

\section*{Methods}
\subsection*{Social dynamics}
Individuals derive  utility $U_M$ from mitigating, and utility $U_N$ from not mitigating. The net utility gained for switching from non-mitigating to mitigating behaviour is therefore $U_M - U_N$. The specific form we assume for the net utility is
\begin{equation}
    U_M - U_N = -(\beta - k) + f(T) + \delta(2x-1) \, .
    \label{net_utility_eqn}
\end{equation}
Here, $(\beta - k)$ represents the net cost associated with switching to mitigating behaviour, $f(T)$ is the cost of climate change impacts, and $\delta$ represents the strength of social norms, which is social pressure that a person feels to conform to prevailing majority opinion. All costs in Eq. \eqref{net_utility_eqn} are perceived costs, so they are informed partly by data, but also have margins built in to account for uncertainty resulting from incomplete information. The rate of change of the proportion of mitigators with time is given by
\begin{equation}
    \frac{dx}{dt} = \kappa x (1-x) [-(\beta - k) + f(T) + \delta (2x-1)] \, . 
    \label{x_eqn_main}
\end{equation}
It is worth noting that this functional form and especially the frequency term $x(1-x)$ could also be taken as a representation of descriptive social norms, since it is through interactions that individuals are able to gauge the average opinion in the population.

We use region-specific empirical estimates for $\beta$, $f(T)$ and $\delta$ in Eq.(\ref{x_eqn_main}), but we need to express these three components in the same utility measure that individuals use to make decisions. We introduce scalings $\hat{\beta}$, $\hat{c}$ and $\hat{\delta}$  so that individuals can react to similar costs in different ways according to their context. This also allows us to reduce the number of free parameters to reduce the chance of overfitting. Unscaled, $\beta$ has dimensions of a cost per person, $f(T)$ of a fraction of income per person, and $\delta$, of a dimensionless quantity between 0 and 1. We divide $\beta$ by per capita income (using SSP 2 forecasts) so this can be interpreted as a fraction of income and then re-write Eq.\ref{x_eqn_main} as follows,

\begin{align}
    \frac{dx}{dt}  &= \kappa x (1-x)[- \hat{\beta} \,\beta(t) + \hat{c} \, f(T) + \hat{\delta} \, \delta (2x-1)] \\
                   &= \kappa \,\hat{\beta}\, x (1-x) \Big[- \beta(t) + \frac{\hat{c}}{\hat{\beta}} \, f(T) + \frac{\hat{\delta}}{\hat{\beta}}  \delta (2x - 1)\Big] \\
                   &= \kappa_0 x (1-x) [-\beta(t) + c_0 \, f(T) + \delta_0 \, \delta (2x-1)],
\end{align}
where $\kappa_0 = \kappa \, \hat{\beta}$, $c_0 = \frac{\hat{c}}{\hat{\beta}}$, and $\delta_0 = \frac{\hat{\delta}}{\hat{\beta}}$.

\subsection*{Coupled model}
The model equations for our coupled social-climate model are
\begin{align}
      \frac{dx_i}{dt} &= \kappa_{0_i} x_i (1-x_i) [- (\beta_i - k)   + c_{0_i}  f_i(T) + \delta_{0_i} \delta_i (2x_i-1)] \label{x_eqn}\\
       \frac{dC_{at}}{dt} &=  \sum_i (1-x_i)\epsilon_i(t) - P + R_{veg} + R_{so} -F_{oc}\label{C_at}\\
    \frac{dC_{oc}}{dt} &= F_{oc}(T,C_{at},C_{oc})\label{C_oc}\\
    \frac{dC_{veg}}{dt} &= P(C_{at},T) - R_{veg}(T,C_{veg}) - L(C_{veg})\label{C_veg}\\
    \frac{dC_{so}}{dt} &= L(C_{veg}) - R_{so}(T,C_{so})\label{C_so}\\
    \frac{dT}{dt} &= \frac{a_E[(F_d(C_{at},T) -\sigma(T+T_0)^4]}{C}\label{T}
\end{align}
where the index $i$ runs over each of the five regions. Eqs.(\ref{C_at})-(\ref{C_so}) describe the the inflow and outflow of carbon in the atmosphere, ocean, vegetation and soil, respectively, and Eq.(\ref{T}) calculates the anomaly in the global mean surface temperature. Details of the functions used in these equations are provided below. 

Eq.(\ref{C_at}) describes the change in atmospheric carbon with time. The first term in the equation represents anthropogenic carbon emissions summed over all regions, with regional emissions $\epsilon_i(t)$,  weighed by the proportion of non-mitigators, $(1-x_i)$.
The emission rate $\epsilon(t)$ represents the worst case annual carbon emissions (in the absence of population-driven mitigation).  Social dynamics begin evolving in our simulations in 2020.  Prior to 2020, $\epsilon(t)$ is fixed according to historical emissions \cite{Guetschow2025}. Thereafter, $\epsilon(t)$ is given by 
\begin{equation}
    \epsilon(t) = \epsilon_{2020} + \frac{(t-2020)\epsilon_{\max}}{(t-2020) + s}
    \label{emissions}
\end{equation}
Here, $\epsilon_{2020}$ represents carbon emissions in 2020, $s$ is the half-saturating constant, and $\epsilon_{\max}$ the limiting value for $\epsilon(t)$, which follows an increasing but saturating trajectory with time (Supplementary Information, Figure \ref{fig:emissions}a).

The second term in Eq.(\ref{C_at}), $P$, represents carbon taken from the atmosphere for photosynthesis, and has the functional form
\begin{equation}
    P(C_{at},T) = \begin{cases} k_pC_{ve0}k_{MM}\Bigg( \frac{p\text{CO}_{2a}-k_c}{K_M+p\text{CO}{2a}-k_c}\Bigg)\Bigg( \frac{(15+T)^2(25-T)}{5625}\Bigg) &\text{when } p\text{CO}_{2a}\ge k_c \text{ and} -15\le T \le2 5
    \\
    0 &\text{otherwise}
    \end{cases}
    \label{photosynthesis}
\end{equation}
Here,  $p\text{CO}_{2a}$ is the ratio of moles of $\text{CO}_2$ in the atmosphere to $k_a$, the total moles of molecules in the atmosphere, and has functional form
\begin{equation}
    p\text{CO}_{2a} = \frac{f_{gtm}(C_{at}+C_{at0})}{k_a},
\end{equation}
where $f_{gtm}$ is the conversion factor used to convert carbon from $Gt$C to moles of carbon. The function in Eq.(\ref{photosynthesis}) is designed to resemble Michaelis-Menten kinetics, with optimal photosynthesis at $T=2$. Eq.(\ref{C_at}) also features the processes of plant respiration, $R_{veg}$, and soil respiration, $R_{so}$, both of which contribute to increasing atmospheric carbon. These are described by functions
\begin{align}
        R_{veg}(T,C_{veg}) &= k_rC_{veg}k_Ae^{-\frac{E_a}{R(T+T_0)}}\\
        R_{so}(T,C_{so}) &= k_{sr}C_{so}k_Be^{-\frac{308.56}{T+T_0-227.13}}
\end{align}
Eq. (\ref{C_oc}) simplifies ocean dynamics, so that the flux of $CO_2$ from the ocean into the atmosphere is given by
\begin{equation}
    F_{oc}(C_{at},C_{oc}) = F_0\chi \Bigg(C_at - \zeta \frac{C_{at0}}{C_{oc0}}C_{oc}\Bigg)
\end{equation}
Here, $\chi$ is the characteristic solubility of $\text{CO}_2$ in water and $\zeta$ is the evasion factor.

Eq.(\ref{C_veg}), in addition to photosynthesis, $P(C_{at},T)$, and plant respiration, $R_{veg}$, accounts for plants releasing carbon into the soil reservoir upon their death, given by
\begin{equation}
    L(C_{veg}) = k_tC_{veg}
\end{equation}

Eq. (\ref{C_so}) describes the uptake of this carbon by the soil as well as the release of carbon through soil respiration, $R_{so}$.  
Finally, Eq. (\ref{T}) assumes a grey-atmosphere approximation, using changes in albedo, $A$, solar flux, $S$, and the opacity of CO$_2$, H$_2$O$_v$ and CH$_4$ to calculate changes in the global mean surface temperature. The net downward flux of radiation absorbed at the earth's surface is
\begin{equation}
    F_d = \frac{(1-A)S}{4}\Bigg(1+\frac{3}{4}\tau\Bigg)
\end{equation}
where $\tau$ is the vertical opacity of the greenhouse atmosphere. The opacities of specific gases are given by
\begin{equation}
    \begin{split}
        \tau(\text{CO}_2) &= 1.73(p\text{CO}_2)^{0.263}\\
        \tau(\text{H}_2\text{O}) &= 0.0126(HP_0e^{-(L/RT)})^{0.503}\\
        \tau(\text{CH}_4) &= 0.0231
    \end{split}
\end{equation}
where $H$ is the relative humidity, $P_0$ is the water vapour saturation constant, $L$ is the latent heat per mole of water, and $R$ is the molar gas constant.  

\subsection*{Parameterization}

\paragraph{Net cost to mitigate ($\beta(t)$):}
 $\beta(t)$ represents the net cost associated with adopting mitigative behaviour. We multiply renewable energy costs from \cite{way2022empirically} with per capita energy demand projections from \cite{bauer2017shared}. We use the per capita costs of the least expensive form of renewable energy available in the market which, historically, has been wind energy. However, projections show solar energy costs falling below those of wind energy around the year 2030 \cite{way2022empirically}; we account for this shift by using the lower of the two costs through time in our calculations. We use costs under a `slow transition' scenario (Supplementary Information, Figure \ref{appendix_beta}a), which assumes wind and solar energy use to grow at rates half their historical rates (costs are assumed decrease with increases in production for a given technology) \cite{way2022empirically}. (All costs are measured as a global average of the Levelized Cost of Electricity (LCOE), in \$[2020]/MWh). To establish region-specific values, we multiply global energy cost forecasts by region-specific per capita energy demand forecasts under the SSP 2 scenario from \cite{bauer2017shared} (Supplementary Information, Figure \ref{appendix_beta}b). From these region-specific costs, we subtract a constant fossil fuel cost, $k$, which is calibrated so that the global average cost of renewables falls below that of fossil fuels in 2035, 2040 and 2060, in the best case, baseline and worst cases; these years are derived from the approximate years in which solar + wind energy provides approximately 50\% of the total electricity generated \cite{way2022empirically}. 

\paragraph{Cost of climate change impacts ($f(T)$):} We use damage estimates from \cite{kotz2024economic} under RCP 8.5 warming, using the provided replication code to generate national damages for 50 bootstraps for a single random seed. From this we calculate the median national damage across all bootstraps, and then use that to calculate population-weighted damage estimates for each region between 2020 and 2100 (using SSP 2 population data from \cite{riahi2017shared}). We use temperature data for RCP 8.5 between 2020 and 2100 from \cite{IPCC2024} to fit a function of the form $f(T) = 1 - a \,\exp (-bT)$ to the regional damage estimates. Fitted coefficients $a$ and $b$ are in Supplementary Information, Table \ref{tab:fT_coeffs}. RCP 8.5 data shows a temperature anomaly of about $1.25^\circ $C in 2020, and our model overestimates this, with a temperature of about $1.43^\circ$ in 2020.

\paragraph{Strength of social norms ($\delta$):} The parameter $\delta$ in our model governs the payoff associated with adhering to social norms. We use survey data from \cite{andre2024globally} to calculate the strength of social norms. Their survey covers 125 countries and was conducted in 2021/2022. We calculate the average of responses on two questions. First, the question that asks if the respondents believe if people on their country `should try to fight global warming', with 1 coding to `yes' and 0 coding to `no, and second, the question asking people about their belief of what fraction of other people in their country are willing to contribute 1\% of their income to fight global warming. We calculate population-weighted average scores for each region in our model (Table \ref{x0_delta_table}).

\paragraph{Initial fraction of mitigators ($x_0$):} We use data from \cite{andre2024globally} to estimate the initial fraction of population supporting mitigation in each region by averaging survey responses on whether people are willing to contribute 1\% of their income to combat global warming (1 codes to `yes' and 0 `codes' to no), and if not, would they be willing to contribute $<1\%$ of their income to combat global warming (0.5 for `yes'), for each country. We then calculate population-weighted average scores for each region in our model (Table \ref{x0_delta_table}).

\paragraph{Emissions ($\epsilon(t)$):} Energy demand data and forecasts \cite{bauer2017shared} allow us to estimate regional proportions of the total energy demand in 2020 and 2100. These proportions are then used to scale $\epsilon_{2020}$ and $\epsilon_{\max}$, respectively, creating region-specific worst case carbon emissions trajectories, $\epsilon_i(t)$ (Supplementary Information, Figure \ref{fig:emissions}b).

\subsection*{Calibration and simulation}
The estimated ranges for $\hat{x}_0$, $\kappa_0$, $c_0$ and $\delta_0$ are chosen to include plausible values based on empirical evidence where available, and also have generous margins beyond these. We allow climate impact costs to range between half of mitigation costs to 15 times mitigation costs (based on supporting evidence \cite{kotz2024economic}), and assume that norms are of the same order of magnitude as mitigation costs. We also multiply the initial condition, $x_0$ by a scaling constant $\hat{x}_0$. Ranges for these parameters are in Supplementary Information, Table \ref{tab:params}. 
 We then use Approximate Bayesian Computation to fit values specific to emissions scenarios, assuming a uniform prior. Distributions of filtered parameters for each scenario are in Supplementary Information, Figure \ref{fig:violin}. Descriptions of and values for climate model parameters used in Eqs. (\ref{C_at}-\ref{T}) are in Table \ref{appendix_climate_parameters}.

Our coupled model was simulated in MATLAB R2021a, using the ode45 module. Social dynamics were initiated in the year 2020, and run until 2200. All empirical estimates we use give us information until 2100; for simulations beyond these, we extrapolate historical functions. 

\section*{Data availability}
Data for the Shared Socioeconomic Pathways is available at \url{https://tntcat.iiasa.ac.at/SspDb}. Historical carbon emissions data is available at \url{https://zenodo.org/records/15016289}. Energy cost forecast data are available upon request to the authors of  \url{https://doi.org/10.1016/j.joule.2022.08.009}. Data on the effects of climate change on income are generated using the code at \url{https://zenodo.org/records/11064757}. Survey data on public support for and perceptions of global warming can be downloaded from \url{https://dataverse.iza.org/file.xhtml?fileId=234&version=5.0}.

\section*{Code availability}
Replication code is available at \url{https://doi.org/10.5281/zenodo.15287833}.
\FloatBarrier
\printbibliography
\FloatBarrier
\clearpage
\appendix
\renewcommand{\thefigure}{S\arabic{figure}} 
\setcounter{figure}{0}

\renewcommand{\thetable}{S\arabic{table}} 
\setcounter{table}{0}
\section*{Supplementary Information}
\subsection*{Data and methods}
\label{appendixA}

\begin{figure}[h]
    \centering
    \includegraphics[width = \linewidth]{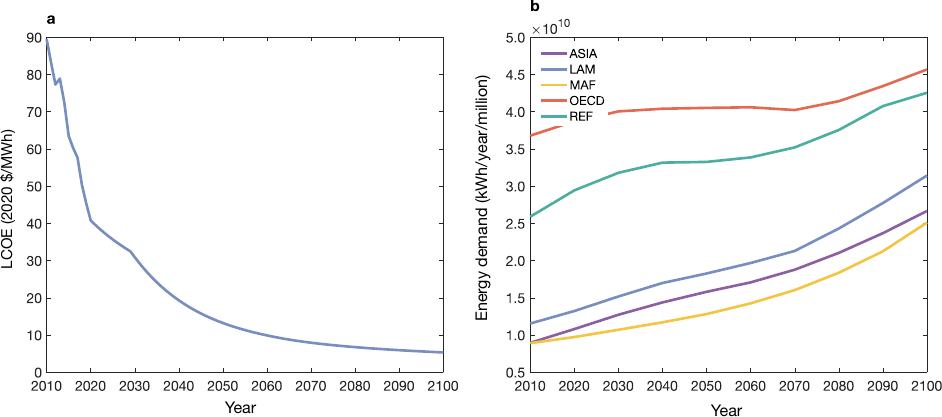}
    \caption{\textbf{Net cost to mitigate:} Shown in (a), costs (minimum of solar and wind energy costs) under the slow transition scenarios (from \cite{way2022empirically}). Historical costs are shown until 2020, beyond which all costs are projections. Solar energy costs fall below wind energy costs around the year 2030. Shown in (b) are energy demand forecasts (in KWh/person (million)/year) for each of the five regions in our model, under SSP 2 (from \cite{bauer2017shared}).}
    \label{appendix_beta}
\end{figure}

\begin{table}[h]
    \centering
    \begin{tabular}{|c|c|c|}
     \hline 
        Region & $a$ & $b$\\
        \hline
        ASIA & 1.49 & 0.30 \\
        LAM &  1.47 & 0.30  \\
        MAF & 1.51  & 0.32   \\
        OECD & 1.25  & 0.17  \\
        REF &  1.14  & 0.09 \\
        \hline
    \end{tabular}
    \caption{\textbf{Fitted coefficients for $f(T)$.} Region-specific fitted coefficients $a$ and $b$ in $f(T) = 1 - a \,\exp (-bT)$, fitting projected damages from \cite{kotz2024economic} to RCP 8.5 temperatures from \cite{IPCC2024}.}
    \label{tab:fT_coeffs}
\end{table}

\begin{figure}
    \centering
    \includegraphics[width=\linewidth]{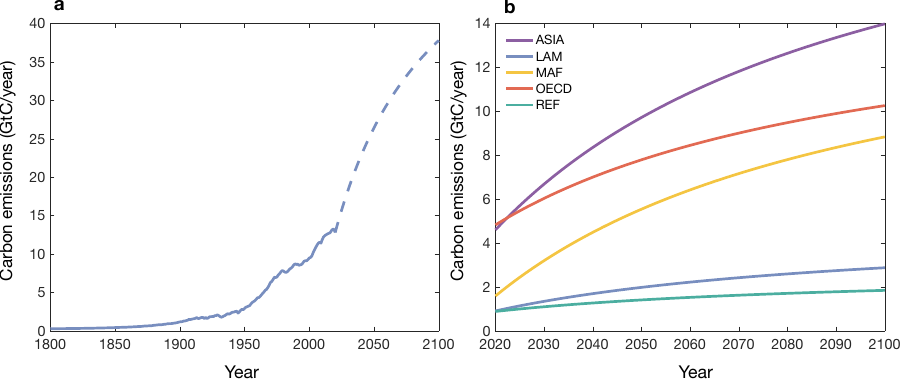}
    \caption{\textbf{Historical and future emissions.} In (a) are historical emissions until 2020 (solid line, based on \cite{Guetschow2025}) and future emissions in the complete absence of mitigation (dashed line), and in (b) are future region-specific emissions when there is no mitigation.}
    \label{fig:emissions}
\end{figure}

\begin{figure}
    \centering
    \includegraphics[width=\linewidth]{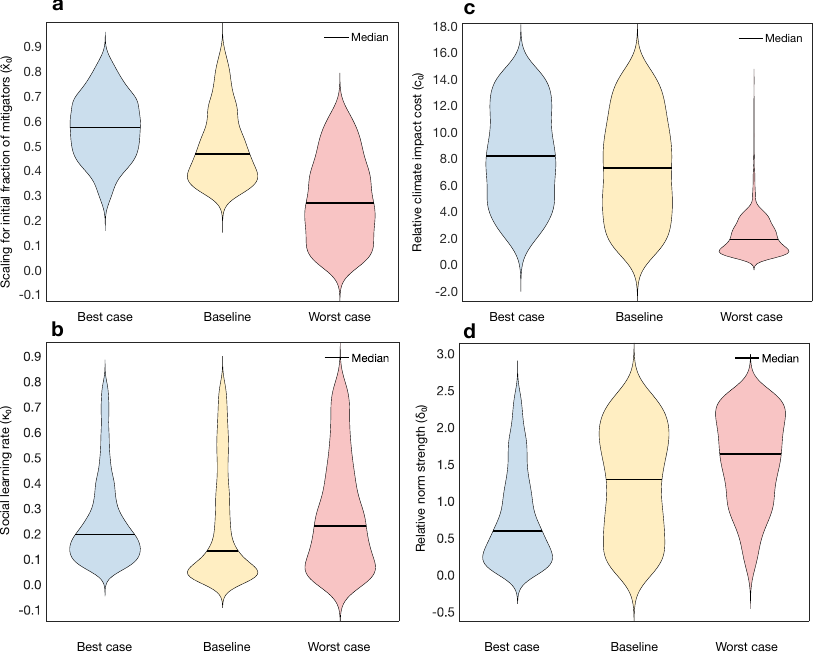}
    \caption{\textbf{Fitted distributions of filtered parameters.} Violin plots show fitted distributions with median values to parameters with 5\% lowest error found using approximate Bayesian computation. (a) shows the scaling for the initial fraction of mitigators ($\hat{x}_0$), (b), the social learning rate ($\kappa_0$), (c), the relative cost of climate impacts ($c_0$), and (d) the relative strength of social norms ($\delta_0$), under best case, baseline and worst case scenarios.}
    \label{fig:violin}
\end{figure}

\begin{table}[ht]
\centering
\resizebox{\textwidth}{!}{ 
\begin{tabular}{|l|c|c|c|c|c|}
\hline
Parameter & Assumed value/range & \multicolumn{3}{c|}{Calibrated value} & Unit \\
\cline{3-5}
                   &                       & Best case & Baseline & Worst case &                \\
\hline
Limiting value for emissions ($\epsilon_{\text{max}}$) &50 & - & - &-  & GtC/year \\
Half-saturation constant for emissions ($s$) & 80 & -& -& -& -\\
Scaling for initial proportion of mitigators ($\hat{x}_0$) & [0.01, 1]    &    0.58   &    0.47     &  0.27  & -   \\
Scaled social learning rate ($\kappa_0$)         & [0.01, 0.8]   &       0.20    &    0.13   &     0.23       & -   \\
Fossil fuel cost ($k$) & - & 435 &359 & 203 & \$/person/year \\
Relative climate impact cost ($c_0$)   & [0.5, 15]    &    8.23  &   7.34   & 1.91     &    -             \\
Relative norm strength ($\delta_0$)  & [0.01, 2.5]           &   0.59    &   1.29     &  1.64    &   -  \\
\hline
\end{tabular}
}
\caption{\textbf{Assumed and calibrated parameter values.} Shown are the assumed ranges/values for each parameter and median filtered values calibrated to RCP emissions under three scenarios using Approximate Bayesian Computation.}
\label{tab:params}
\end{table}

\begin{table}[h]
    \centering
    \begin{tabular}{c|c|c|c}
        Parameter & Definition & Baseline values& Unit  \\
        \hline
        C$_{at0}$ & initial CO$_2$ in atmosphere & 596 & GtC \\
        C$_{ao0}$ & initial CO$_2$ in ocean reservoir & 1.5 $\times$ $10^5$ & GtC  \\
        C$_{veg0}$ & initial CO$_2$ in vegetation reservoir & 550 & GtC \\
        C$_{so0}$ & initial CO$_2$ in soil reservoir & 1500 & GtC \\
        $T_0$ & initial average atmospheric temperature & 288.15 & K \\
        $k_p$ & photosynthesis rate constant & 0.184 & yr$^{-1}$ \\
        $k_{MM}$ & photosynthesis normalizing constant & 1.478 & 1 \\
        $k_c$ & photosynthesis compensation point & $29 \times 10^{-6}$ & 1 \\
        $K_M$ & half-saturation point for photosynthesis & $120 \times 10^{-6}$ & 1 \\
        $k_a$ & mole volume of atmosphere & $1.773 \times 10^{20}$ & moles \\
        $k_r$ & plant respiration constant & 0.092 & yr$^{-1}$ \\
        $k_A$ & plant respiration normalizing constant & 8.7039 $\times 10^9$ & 1 \\
        $E_a$ & plant respiration activation energy & 54.83 & J mol$^{-1}$\\
        k$_{sr}$ & soil respiration rate constant & 0.034 & yr$^{-1}$ \\
        k$_B$ & soil respiration normalizing constant & 157.072 & 1 \\
        k$_t$ & turnover rate constant & 0.092 & yr$^{-1}$ \\
        $C$ & specific heat capacity of Earth's surface & 4.69 $\times 10^{23}$ & J K$^{-1}$ \\
        $a_E$ & Earth's surface area & 5.101 $\times 10^{14}$ & m$^2$ \\
        $\sigma$ & Stefan-Boltzmann constant & 5.67$\times 10^{-8}$ & W $m^{-2}K^{-4}$ \\
        $L$ & latent heat per mole of water & 43.655 & mol$^{-1}$ \\
        $R$ & molar gas constant & 8.314 & J mol$^{-1}$ K $^{-1}$ \\
        $H$ & relative humidity & 0.5915 & 1 \\
        $A$ & surface albedo & 0.225 & yr$^{-1}$ \\
        $S$ & solar flux & 1368 & Wm$^{-2}$ \\
        $\tau(\text{CH}_4)$ & methane opacity & 0.0231 & 1 \\
        $P_0$ & water vapor saturation constant & 1.4 $\times 10^{11}$ & Pa \\
        $F_0$ & ocean flux rate constant & 2.5 $\times 10^{-2}$ & yr$^{-1}$ \\
        $\chi$ & characteristic CO$_2$ solubility & 0.3 & 1 \\
        $\zeta$ & evasion factor & 50 & 1\\
    \end{tabular}
    \caption{\textbf{Climate model parameters:} Parameter values used in the climate component of the model; all values are taken from \cite{bury2019charting}.}
    \label{appendix_climate_parameters}
\end{table}

\FloatBarrier
\subsection*{Results}
\label{appendixB}
\begin{figure}[h!]
    \centering
    \includegraphics[width=0.9\linewidth]{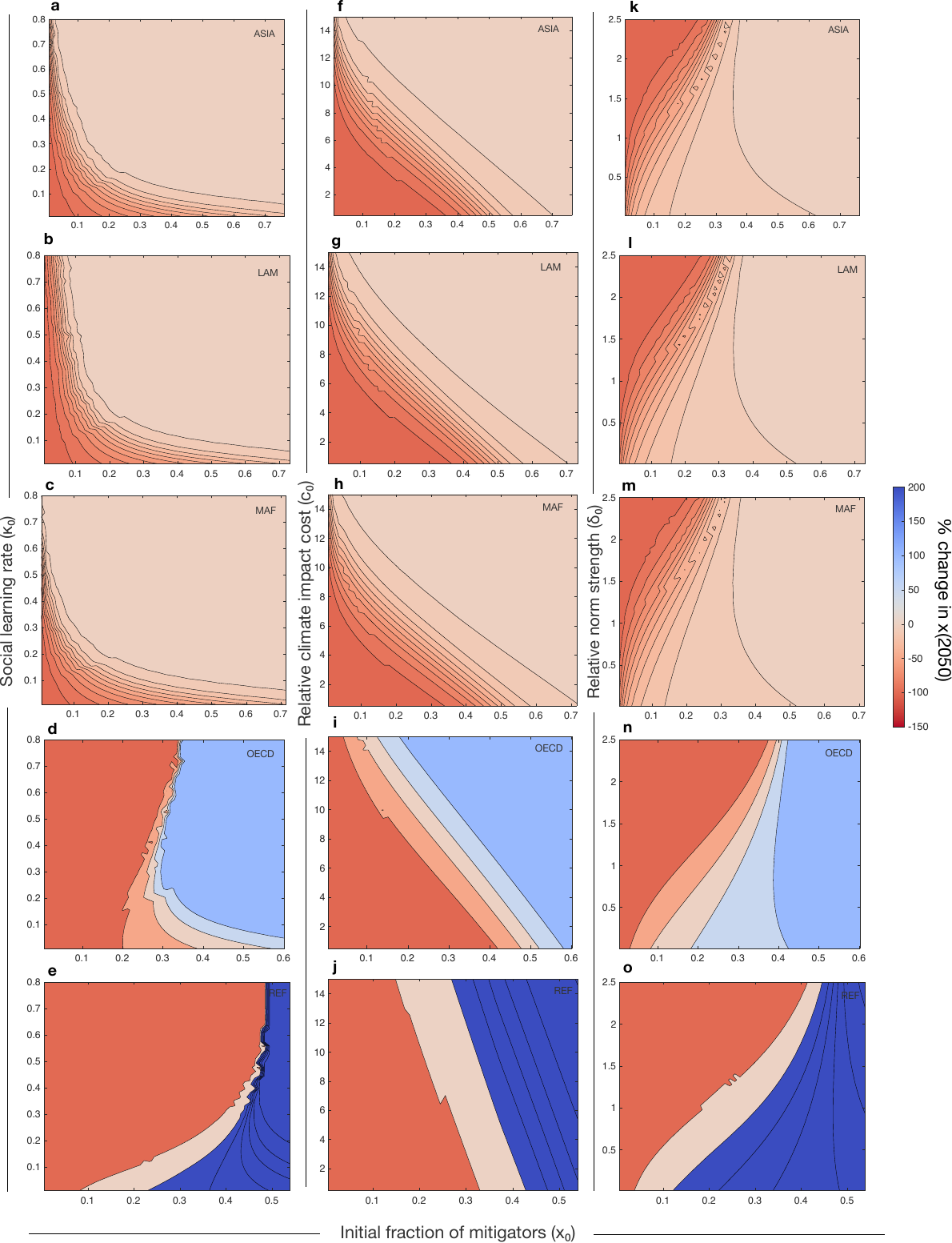}
    \caption{\textbf{Strong initial support helps mitigation.} Shown is the percentage change in the fraction mitigation supporters ($x_0$) in each region in 2050, relative to baseline, for various parameters being changed in conjunction with the initial fraction of supporters.}
    \label{fig:SI_param_planes_4_6}
\end{figure}

\begin{figure}
    \centering
    \includegraphics[width=\linewidth]{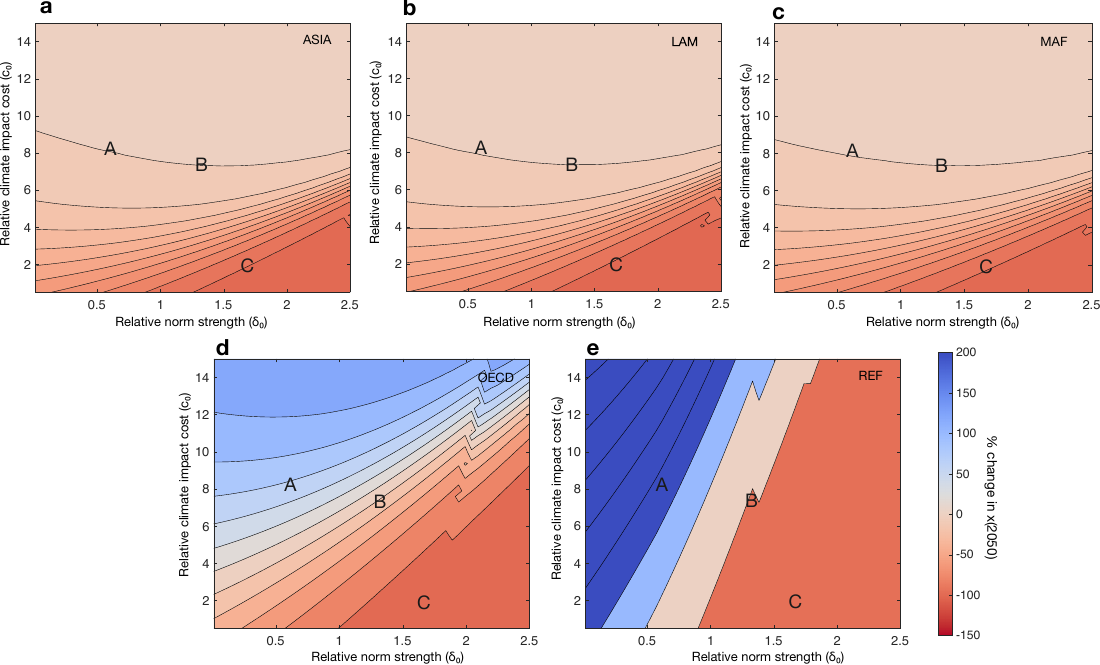}
    \caption{\textbf{Increasing the relative climate impact cost has diminishing returns, unlike relative norm strength.} Shown is the percentage change in fraction of mitigation supporters in 2050, relative to baseline, for each region for various combinations of the relative climate impact cost ($c_0$) and relative norm strength ($\delta_0$). On each plane, `A' marks the best case, `B', baseline, and `C', the worst case parameter combinations. }
    \label{fig:param_planes_3_SI}
\end{figure}

\begin{figure}
    \centering
    \includegraphics[width=\linewidth]{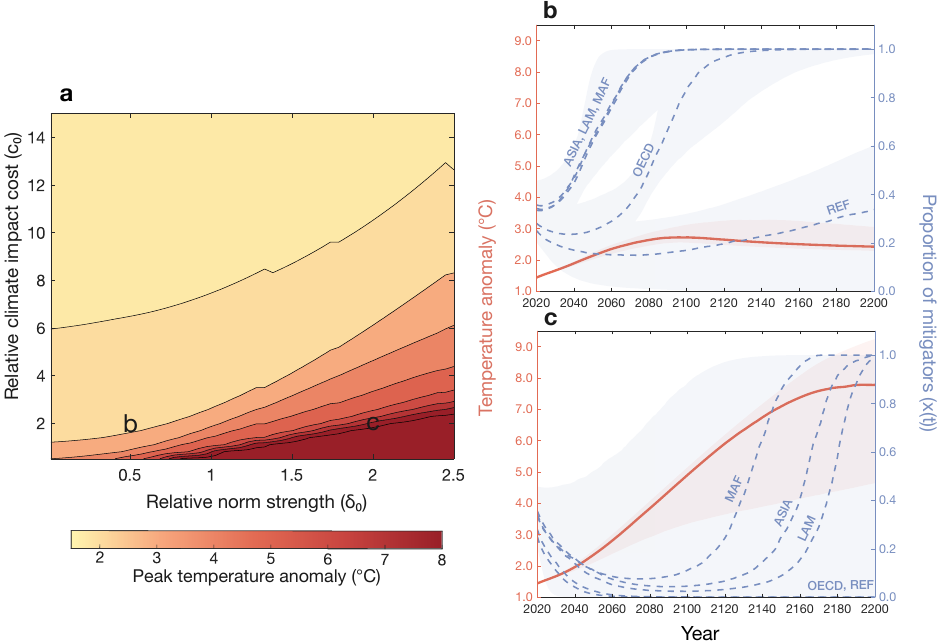}
    \caption{\textbf{Strong norms coupled with low relative climate impact costs lead to severely high temperatures.} (a) shows the peak temperature anomaly recorded for various combinations of  the relative climate impact cost ($c_0$)  and relative norm strength ($\delta_0)$ varied for all regions simultaneously. The time evolution of temperature and mitigation support for marked points `b' and `c' are shown in (b) and (c), respectively, with temperature in red (left y-axis) and mitigation support in blue (right y-axis).}
    \label{fig:norms_temp}
\end{figure}

\begin{figure}
    \centering
    \includegraphics[width=\linewidth]{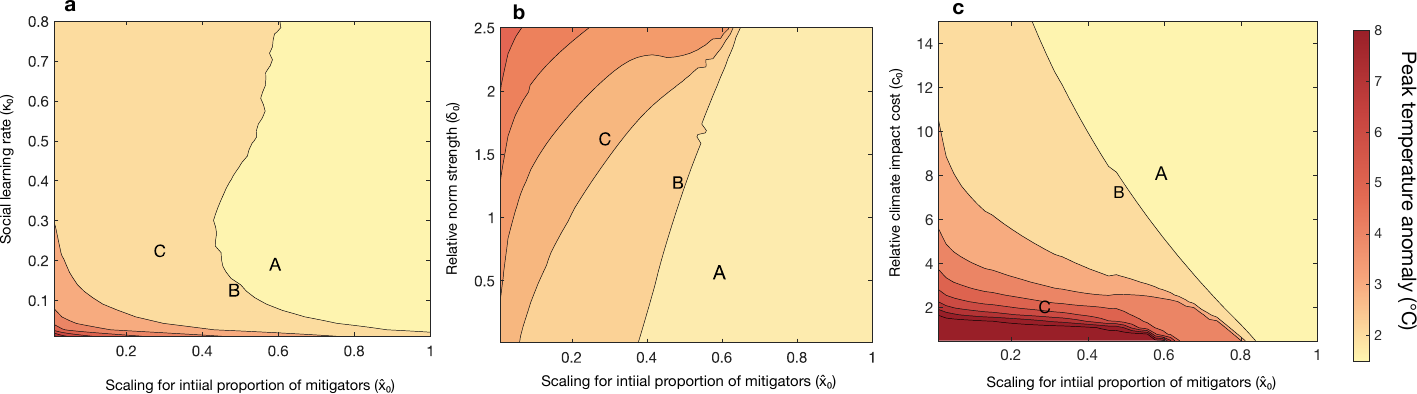}
    \caption{\textbf{Weak initial mitigation leads to higher peak temperatures.} (a)-(d) show the peak temperature anomaly recorded for various parameter combinations for all regions together. On each plane, `A' marks the best case, `B', baseline, and `C', the worst case calibrated parameter combinations.}
    \label{fig:placeholder}
\end{figure}
\end{document}